# Structure and Superconductivity in Zr-Stabilized, Nonstoichiometric Molybdenum Diboride

L. E. Muzzy[1], M. Avdeev[2], G. Lawes[3], M.K. Haas[1], H.W. Zandbergen[1,4], A.P. Ramirez[3], J.D. Jorgensen[2], and R.J. Cava[1]

[1]Department of Chemistry and Princeton Materials Institute, Princeton University, Princeton NJ 08544
[2]Materials Science Division, Argonne National Laboratory, Argonne IL 60439
[3]Los Alamos National Laboratory, Los Alamos NM 87545
[4]National Centre for HREM, Laboratory of Materials Science, Delft University of Technology, Rotterdamseweg 137, 2628 AL Delft, The Netherlands

**Abstract**

The structure and physical properties of the Zr-stabilized, nonstoichiometric molybdenum diboride superconductor are reported. Good quality material of the diboride structure type can only be obtained by partial substitution of Zr for Mo, and the quenching of melts. The diboride phase is best made with boron in excess of the ideal 2:1 boron to metal ratio. Powder neutron diffraction measurements show that the non-stoichiometry is accommodated by atom deficiency in the metal layers. The diboride structure type exists for $(Mo_{.96}Zr_{.04})_xB_2$ for $1.0 \geq x \geq 0.85$. Electron diffraction shows that the stoichiometric material, x=1, has a significant number of stacking faults. $T_c$ increases from 5.9 to 8.2K with the introduction of metal vacancies. Resistivity measurements indicate that $(Mo_{.96}Zr_{.04})_{.88}B_2$ is a bad metal, and specific heat measurements show its electronic density of states is $\gamma = 4.4$ mJ/mol $K^2$, and that $\Delta C/\gamma T_c = 1.19$. Preliminary boron isotope effect measurements indicate an exponent $\alpha = 0.11 \pm 0.05$. Analysis of the data in terms of the electronic structure is reported, allowing an estimate of the electron-phonon coupling constant, $\lambda \approx 0.1$-$0.3$, making these weak-coupling superconductors. Preliminary characterizations of the superconductivity in the related phases $Nb_xB_2$ and $(Mo_{.96}X_{.04})_{.85}B_2$ for X=Ti, and Hf are reported.

## Introduction

The recent discovery of superconductivity in $MgB_2$ at 39K (1) has re-kindled interest in the possibility of attaining high $T_c$s in conventional intermetallic compounds. Of general interest are compounds where the presence of light atoms might give rise to high frequency phonons coupled to the conduction electrons, and of more specific interest are superconductors where boron plays an essential structural role. The crystal structure of $MgB_2$ is that of the now familiar $AlB_2$ ("diboride") type, with graphite-like honeycomb-geometry layers of boron separated by a hexagonal layer of Mg. The chemical and structural aspects of $MgB_2$ have proven to be more complex than might be envisioned from its simple formula. In the time since the discovery of $MgB_2$, reconsideration of other possible diboride-type superconductors has indicated that the original reports of superconductivity in the 1970s in "diborides" based on Mo are indeed correct, with others, such as $TaB_2$ and $ZrB_2$, not superconducting (2).

The early transition metal diborides were initially tested for superconductivity to clarify emerging ideas about the optimal electron/atom (e/a) ratio for superconductivity in intermetallic compounds. Based upon this work, correlating the maximum Tc for many intermetallic compounds with e/a ratios of 5-7, Y, Zr, Nb and Mo diborides were studied, and the latter two were made superconducting by appropriate doping and preparation (2). It was reported that although stoichiometric $NbB_2$ is not superconducting, adding excess boron, to an optimal composition of $NbB_{2.5}$, led to superconductivity at 3.9 K. In addition, superconducting transition temperatures were reported to increase somewhat by partially substituting other transition metals for the Nb.

The case of "molybdenum diboride' was the most interesting, and motivated the present study. $MoB_2$ only exists at temperatures above 1500 C in the Mo-B system (3), and cannot be retained to room temperature without special quenching methods. Materials of that nominal composition were not found to be superconducting. However, forcing excess boron into the lattice by splat-melting a composition $MoB_{2.5}$ was found to yield a material with the $AlB_2$ structure type and a superconductor with a transition temperature of 7.45 K (2). The thermodynamically stable phase at composition $MoB_{2.5}$ ($Mo_2B_5$) has a different crystal structure and is not superconducting above 2 K (4). As was the case with Nb, partial substitution of Mo by other metals was reported to produce more stable superconductors. In particular, doping with zirconium was reported to increase the superconducting onset temperature to near 11 K.

Here we report structural and superconducting property studies of Zr-stabilized $AlB_2$ type moldbdenum diboride over a significant range of boron to metal ratios. We find that the nonstoichiometry is accommodated by the formation of metal site vacancies. We find that the $AlB_2$ type phase is thermodynamically unstable in this chemical system at temperatures of 1200 degrees and below. We have not seen evidence for $T_c$s above 8.3 K. Our preliminary boron isotope effect measurements indicate a small dependence of Tc on boron mass, implying relatively smaller contribution of boron states to the superconductivity when compared to $MgB_2$. Comparison of the observed and calculated density of states at the Fermi energy suggests that $MoB_2$ is a weak coupling superconductor.

**Synthesis**

Initially, a general synthetic study of $Mo_{2-x}Zr_xB_{5-y}$ and $Nb_{2-x}Zr_xB_{5-y}$ compositions was undertaken. Half gram samples were arc melted under argon (three times). Mo foil, Zr foil, niobium powder and lump boron were employed as starting materials. Powder X-ray diffraction (Copper Ka radiation) was employed to determine phase purity. Initial studies confirmed the approximate stoichiometry ranges initially reported. However, on employing neutron diffraction to investigate samples that appeared to be single phase by X-ray diffraction, the former method being proportionally much more sensitive to light elements, it was found that only through the use of neutron diffraction could the true phase purity of samples be determined. During the course of the general studies it was found that annealing as-melted samples at temperatures between 900 and 1200 C always resulted in the disappearance of the $AlB_2$ type phase in the Mo-Zr-B system, consistent with the fact that the $AlB_2$-type phase in the Mo-B system appears to be thermodynamically stable only at very high temperatures. The Zr-substitution apparently suppresses the kinetics of the phase decomposition on cooling, though it doesn't extend the thermal stability region to temperatures as low as 1200 C. The result is that the $AlB_2$ phase can be retained as a metastable phase during the cooling process that takes place in the arc melter in the materials containing small amounts of Zr. This data showed that what appeared to be a single phase material of the $AlB_2$ type could be obtained only for approximately 4% substitution of Zr for Mo, and B contents between approximately 2.0 – 2.5 per metal. A final set of samples, of stoichiometry $(Mo_{.96}Zr_{.04})B_y$, synthesized as above, was made with $^{11}B$ in the stoichiometry range $2.0 < y < 3.1$. Approximately 0.5 gram pieces of the samples with clean, sharp X-ray diffraction patterns that appeared to be those of single phase compounds were then analyzed by neutron diffraction at IPNS. For the isotope effect studies, a set of comparable samples was made with $^{10}B$. For the electron microscope study, samples were made at selected compositions by the arc-melting method as described above. Finally, a variety of $AlB_2$-type samples in the Nb-Zr-B and Mo-M-B systems were prepared with natural abundance boron to test their superconducting properties.

**Crystal Structure**

Neutron powder diffraction data were collected on the Special Environment Powder Diffractometer at the IPNS at Argonne National Laboratory. As-melted samples of $(Mo_{0.96}Zr_{0.4})B_y$ with the boron content, y, varying from 2.0 to 3.1 were measured. Samples were finely ground before data collection to avoid the influence of preferred orientation and contained in thin-walled vanadium cans. The average data collection time was 16 hours per sample. Depending on the starting composition, samples were found to contain the hexagonal diboride phase, and, in some cases, elemental boron as a second phase.

The structural refinement of the neutron diffraction data indicated that when excess boron is incorporated in the diboride phase, it is accommodated by creating vacancies on the metal (Mo,Zr) site rather than by excess boron going into the lattice interstitially. The expectation that excess boron could be accommodated at interstitial sites comes from literature regarding the closely related $Mo_2B_{5-x}$ phase. An early report of this phase claimed (without experimental proof) that the structure contained both honeycomb B planes as are found in $AlB_2$ type materials and additional layers of planes in which the hexagonal boron rings were capped by an additional (i.e. interstitial) boron (4). Later work, using X-ray single-crystal diffraction (5) confirmed the existence of the puckered layer, but did not locate the interstitial boron atoms. A subsequent study using neutron powder diffraction detected the interstitial boron sites, but with a small occupancy (6). In a previous study of $MoB_{2+x}$, X-ray diffraction and density measurements were compared to conclude that the excess boron was in interstitial sites (7).

Ordering of the metal vacancies in the $Mo_{1-x}B_2$ structure is possible, but no superlattice peaks that might arise from such ordering were seen in the neutron diffraction data. Electron diffraction (described below) is far more sensitive to ordering of this type as it has a much larger dynamic range for observation of weak peaks, and also is sensitive to ordering on short length scales. Very weak superlattice peaks were in fact observed by electron diffraction, as described below, but their origin is not clear at this time.

Reitveld refinements using the neutron diffraction data provide a description of the structural and chemical characteristics of the samples. A representative diffraction pattern is presented in figure 1. For samples with boron to metal ratio greater than 2.2:1, two phases are present, the $AlB_2$ type molybdenum zirconium diboride, and elemental boron. The second phase present in the samples with nominal boron to metal ratios greater than 2.3:1 was identified as rhombohedral beta boron. When refining all diffraction data, the structural parameters of the second phase elemental boron were fixed in accordance with the results of the single crystal structure determination (8), with its phase fraction and cell parameters being the only variables. The pattern shown in figure 1 is from a high boron content sample, chosen to illustrate the presence of the two phases in some samples.

The refined structural and chemical information for samples with nominal boron to metal ratios 2:1 to 2.9:1 is presented in table 1. Data for samples with boron to metal rations in excess of 2.9:1 are not presented because no new information is gained through analysis of those materials, as described below. Primary structural parameters determined were the lattice parameters of the diboride phase, the fractional concentration of metal atoms on the metal site in the diboride phase, and the wt% of the second phase of elemental boron. The fraction of metal site occupancy in the diboride phase as a function of the nominal boron to metal ratio in the as-prepared samples, determined from the structural refinements, is shown in figure 2. The data show that no additional boron enters the diboride phase beyond a nominal boron to metal ratio of 2.4-2.5:1. At a B:M ratio of 2:1, a stoichiometric diboride is formed, of composition $(Mo_{.96}Zr_{.04})B_2$. Thus, surprisingly, the small amount of zirconium added has allowed for the metastable retention of the diboride type phase to room temperature for the cooling rates attained by arc-melting. The electron microscope images described below show that this stoichiometric diboride phase has a significant number of $c$-axis stacking defects. As the proportional amount of boron is increased, metal atoms are removed from the structure to a limit of about 15% vacancies on the metal sites before no more vacant sites are possible for these synthesis conditions. Thus the stoichiometry range of the diboride phase we find is $(Mo_{.96}Zr_{.04})_xB_2$, $1.00 \geq x \geq 0.85$. The microscopy images show that the overall crystallinity of the metal deficient phase is much improved over that of the stoichiometric compound.

Boron begins appearing as a second phase at B/M ratios greater than 2.2:1. As figure 2 shows, however, the composition of the diboride phase continues to change to B:M ratios up to approximately 2.5:1, even as the second phase of boron is increasing in relative proportion. Since this violates the Gibbs phase rule for a two-component system, we conclude that thermodynamic equilibrium is not achieved during the synthesis. This is of course not surprising for synthesis by arc melting without annealing, which we find to be necessary to make the phase, but it does imply that different results (a different range of metal vacancy concentrations in the diboride phase, for example) might be achieved with different synthesis techniques.

As a check of the validity of the results, the sum of the amount of experimentally determined second-phase boron and the boron in the diboride phase can be compared to the nominal (starting) amount of boron. The results of this comparison are shown in figure 3. The agreement shows that all boron is accounted for and confirms the accuracy of the structural analysis technique. Moreover, the agreement shows that the metal-site vacancy model is correct and accounts for all of the "excess" boron in the diboride phase. The small deviations are perhaps the best indication of the overall error bars, which are likely a combination of statistical error bars from the refinement, and errors in composition arising during the synthesis.

The variation of the lattice parameters of the diboride phase as a function of metal atom stoichiometry is presented in figure 4. Starting from the stoichiometric composition, the in-plane lattice parameter shrinks, as expected on removing atoms, though the stiffness of the boron net restricts the shrinkage to a modest value. The out-of-plane cell parameter increases as the metal atoms are removed. This suggests a weakening of interplane bonding with increasing metal deficiency.

Electron microscopy was performed on $Mo_{1-x}Zr_xB_y$ samples with nominal compositions x=0.025 and 0.05, and y = 2.0, 2.2 and 2.5. Electron transparent areas were obtained by crushing, suspending in ethanol, and placing a few droplets on a Cu grid with a holey C film. Electron diffraction studies were performed with a Philips CM300UT-FEG and a Philips CM200ST-FEG electron microscope, both with a field emission gun. Nanodiffraction was performed using a condenser aperture of 10 μm and an electron probe size of 4-10 nm in diameter. High resolution electron microscopy (HREM) was performed with the Philips CM300UT-FEG electron microscope.

Electron diffraction study of the specimens with stoichiometric boron content, $(Mo_{.975}Zr_{.025})B_2$ and $(Mo_{.95}Zr_{.05})B_2$, show strong streaking along the c*direction, as is shown in figure 5. Some of the crystallites in the boron excess samples also show a similar streaking, although when present it is weaker in intensity. Electron microscopy of the $(Mo_{.975}Zr_{.025})B_2$ sample shows (001) planar defects and strain associated with these defects, as can be seen in figures 6 a and b. The planar defects occur throughout the crystals in this stoichiometric material, with a density varying from 50 to 300 per micrometer. HREM images (figure 6b and inset) show that the planar defects have a structure with a *c* axis longer than that of the average structure (*c*=3.12 angstroms) by about 1 angstrom. Across the defect a lattice image shift can be observed, which - in combination with the elongation along the *c* axis - suggests that the defect structure is related to that of the diboride by the inclusion of an extra $Mo_{1-x}Zr_x$ plane. More detailed HREM analysis would be needed to verify this model for the defects, but the fact that they are metal-excess defects is fully consistent with their disappearance in compositions which are metal deficient. Assuming an average defect spacing of 6 nm, the presence of a double $Mo_{1-x}Zr_x$ layer at the defect, and fully occupied B planes, the composition of the defect phase would be $(Mo_{1-x}Zr_x)_{1.05}B_2$. Nanodiffraction with an electron beam diameter of 3-4 nm shows that the areas between the defects do not contain any streaking and are thus defect-free. Interestingly, to add up to the stoichiometric boron to metal ratio, the above analysis suggests that the defect-free regions are metal deficient, consistent with our observations that the overall stability of the diboride phase in this system is increased for materials in which the boron to metal ratio is larger than 2:1.

Figure 7 shows a $[011]_b$ diffraction pattern (b denotes the basic unit cell) taken for a sample in which the diboride phase has composition $(Mo_{.96}Zr_{.04})_{.85}B_2$. It shows the presence of weak, diffuse superlattice reflections among the reflections of the $AlB_2$-type parent structure. These reflections are at the positions $h/3,k/3,l/2$, giving an $a\sqrt{3},a\sqrt{3},2c$ superstructure. (The superreflections are actually not located in this diffraction plane in figure 7, they can still be seen because they are very broad.) Because the superreflections are very weak, they do not appear to be associated with ordering of the metal atoms and vacant metal atom sites. Preliminary investigation suggests that they do scale in intensity with the Zr substitution content of the phase, but that they remain very weak, indicating that the structural modification they represent is not very pronounced. Further study would be necessary to clarify the origin of this superlattice.

**Characterization of Superconductivity**

The superconductivity was characterized for the same $^{11}$B-based samples employed in the neutron diffraction study. The general nature of the superconducting transitions was characterized by measuring the temperature dependent magnetizations for samples in a commercial magnetometer in an applied DC field of 15 Oe. Magnetizations were measured on heating after cooling in zero field. The results are shown in figure 8. The curves are labeled with the experimentally determined composition of the diboride phase in the sample, spanning the range of allowed metal atom stoichiometries. The figure shows that the diboride phase is a bulk superconductor at all compositions (differences in the magnitude of the observed screening below Tc are due to experimental variations). The stoichiometric material, $(Mo_{.96}Zr_{.04})B_2$, displays the lowest superconducting transition temperature and the broadest superconducting transition. The relative breadth of the transition is due to the structural disorder and accompanying strong lattice strain observed in the electron microscopy study. The nonstoichiometric phases display sharper transitions, consistent with the reduction in structural disorder observed by microscopy.

The superconducting transition temperature increases as the metal atom site nonstoichiometry increases: from a value of 5.9K for $(Mo_{.96}Zr_{.04})B_2$ to a value of 8.2K for $(Mo_{.96}Zr_{.04})_{.85}B_2$. The transition temperature clearly saturates to near its maximum value after the introduction of 5-7 % metal site vacancies, as shown in figure 9. The potential effect of metal site nonstoichiometry in this system is complex and would not be expected to be the same as only removing electrons from the stoichiometric compound: the degree of structural disorder is decreased in terms removing intergrowth layers but increased in terms of putting vacant sites in large proportion on the electronically active sites. Thus the microscopic reason for the observed variation in Tc is not obvious, but the data suggest that the stacking defects and associated strain are much more detrimental to the superconductivity than random metal site vacancies. The original report (2) suggested a relationship of Tc to the ratio of the *c*-axis and *a*-axis crystallographic cell parameters. Our results for this correlation are shown in the inset to figure 9. Although we observe the same general behavior as previously reported, Tc saturates at a considerably lower Tc than the "onset temperatures" previously reported (2). In fact, for approximately 50 samples we synthesized throughout the Mo-Zr-B system by arc melting, both annealed and as-cooled, we saw no sign of superconductivity, even in small amounts, at temperatures greater than 8.5K. Higher Tc materials in this system, therefore, if they exist, may require more extreme synthetic methods such as "splat cooling" (2).

The electrical resistivity was characterized in the standard 4-probe configuration, with leads attached by silver epoxy. The temperature dependent resistivity for a sample with superconducting phase composition $(Mo_{.96}Zr_{.04})_{.85}B_2$ is shown in figure 10.

This sample is approximately 98 mol % diboride phase, so the second phase boron present is not expected to significantly effect the measured resistivity. The diboride superconductor is shown to be a poor metal, with ambient temperature resistivity of about 1.4 mohm-cm, Only a very small decrease of resistivity is observed on cooling, dropping only to 1.2 mohm-cm at 10K. The characteristics of the resistivity are consistent with disorder scattering introduced by the presence of 12% vacancies on the metal atom site. The inset to figure 10 shows a detail of the resistivity transition. The onset is approximately 8.5K, and the 90-10% width is approximately 0.05K.

The specific heats in the vicinity of the superconducting transition were measured by the adiabatic heat pulse method in a commercial apparatus. The results are shown in figure 11, for diboride phases of composition $(Mo_{.96}Zr_{.04})B_2$ and $(Mo_{.96}Zr_{.04})_{.88}B_2$. The former sample is single phase and the latter is 98 mol % diboride phase and 2% elemental boron. The upper panel shows the variation of C/T vs. T. Using the equivalent area method, the superconducting transition is characterized by $\Delta C/\gamma Tc=1.19$ for $(Mo_{.96}Zr_{.04})_{.88}B_2$. The value for the stoichiometric compound $(Mo_{.96}Zr_{.04})B_2$, $\Delta C/\gamma Tc=1.8$, we consider much less reliable due to the rounding of the superconducting transition caused by the defective nature of this composition. The value $\Delta C/\gamma Tc=1.19$ for $(Mo_{.96}Zr_{.04})_{.88}B_2$ is consistent with that expected for a BCS superconductor, where the ideal value is 1.43, and very similar to that observed for $MgB_2$, 1.09 (9). The lower panel shows the variation of C/T with $T^2$, employed to determine the electronic densities of states at the Fermi energy.

The data show that the electronic density of states is somewhat larger ($\gamma$= 4.4 mJ/mol K$^2$) for the higher Tc material (Mo$_{.96}$Zr$_{.04}$)$_{.88}$B$_2$ than it is for the lower Tc stoichiometric material ($\gamma$ = 3.65 mJ/mol K$^2$) (Mo$_{.96}$Zr$_{.04}$)B$_2$.

Preliminary boron isotope effect measurements were performed for (Mo$_{.96}$Zr$_{.04}$)$_{.88}$B$_2$ for samples made with $^{10}$B and $^{11}$B. This composition was selectred for study due to the high reproducibility of observed Tcs. The superconducting transitions were characterized by specific heat measurements. A detail of the transition regions is shown in figure 12. The data show that the transition in the $^{11}$B sample occurs at 8.25±0.02 K, and in the $^{10}$B sample at 8.34±0.02 K. (Transition temperatures were determined by extrapolation of line of the steeply rising heat capacity down to the extrapolation of the normal state baseline heat capacity above Tc). The boron isotope effect exponent, in Tc $\propto$ M$^{-\alpha}$, is estimated from these data to be $\alpha$ = 0.11 ± 0.05. This value is less than that determined for boron in MgB$_2$, $\alpha$ = 0.26–0.30 (10,11), indicating that the contribution of boron states to the superconductivity is considerably less in molybdenum diboride superconductors than in MgB$_2$. We note that the Debye temperatures estimated from the data in figure 12 indicate $\Theta_D$ = 620 ± 30K for (Mo$_{.96}$Zr$_{.04}$)$_{.88}$$^{11}$B$_2$ and $\Theta_D$ = 690 ± 30K for (Mo$_{.96}$Zr$_{.04}$)$_{.88}$$^{10}$B$_2$, consistent with the lower mass of the latter material.

Finally, we show in figures 13 and 14 the magnetic characterization of the superconducting transitions in related diboride materials based on both Mo and Nb (materials synthesized with natural abundance boron). Figure 13 shows the magnetic characterization of materials of composition (Mo$_{.96}$X$_{.04}$)$_{.8}$B$_2$, found by X-ray diffraction to be dominated by material of the diboride type, though not single phase. Interestingly, though all metal atom substitutions do result in the stabilization of the diboride structure type, not all substitutions lead to superconductivity above 4 K. No transition was observed for Ta and W dopants above 4 K. The onset of the transition for the Ti-substituted material is approximately 5 K, and only the Hf-substituted material shows a transition temperature comparable to that of the Zr substituted material. It may be of future interest to determine the reason for the differences in the Tcs of these highly related compounds. Figure 14 shows the transitions for superconducting materials based on NbB$_2$, which is not superconducting above 2 K at its stoichiometric composition. The introduction of niobium deficiency, however, results in the appearance of superconductivity with a Tc of approximately 3.5K. The compound of stoichiometry Nb$_{.9}$B$_2$, whose superconducting transition is shown in figure 15, appears single phase by X-ray diffraction, consistent with the reported phase equilibrium diagram (12). Mo and Zr-substituted materials, deficient in metals, also appeared single phase in X-ray diffraction characterization, though they may contain undetected elemental boron, and have very similar Tcs to the nonstoichiometric niobium diboride. The transitions appear to be broader than that of Nb$_{.9}$B$_2$, suggesting the presence of inhomogeneities. We note, finally, that similarly substituted and nonstiochiometric materials based on the non-superconducting pure diboride TaB$_2$ (i.e. (Ta$_{.9}$Zr$_{.1}$)$_{.8}$B$_2$) can be made with very good quality by arc melting. We have found no superconducting transitions in the Ta-based diborides at temperatures above 2 K.

**Electronic Structure**

In order to gain more insight into the superconducting behavior of these materials, the electronic structure of stoichiometric $MoB_2$ was calculated. The experimental lattice parameters of $Mo_{.96}Zr_{.04}B_2$, a=3.055 Å and c=3.128 Å, in the hexagonal space group P6/mmm were employed. Calculations were performed using the self-consistent TB-LMTO-ASA codes of Anderson et. al. (13) within the local density approximation. Boron 2$s$, 2$p$, 3$d$ and Molybdenum 5$s$, 5$p$, 4$d$ comprised the valance basis set with Boron 3$d$ orbitals handled by a special down folding procedure (14). Integration was performed using the tetrahedron method (15). The number of $k$-points used in the calculation was incrementally increased until there was no change in the total energy, with a final total of 793 irreducible $k$-points from a grid of 22 reducible points in each direction of the Brillion zone.

Figure 15 plots the density of states versus energy for $MoB_2$. The general results are in good agreement with those previously reported for $MoB_2$ (16) and other transition metal diborides (16,17). The partial density of states for B and Mo are delineated, illustrating that the states at the Fermi level are derived primarily from molybdenum. They are approximately 75% Mo $d$ in character. This is in contrast to the isostructural superconductor $MgB_2$ (17,18), in which states at the Fermi energy are derived approximately 60% from the Boron sublattice. In addition, in $MgB_2$ the Fermi energy is dominated by Boron sigma bonding interactions, while in $MoB_2$ the Boron sigma bonding orbitals are completely filled and the Fermi energy is dominated by Mo-Mo antibonding interactions. The Boron states that are present at the Fermi energy in $MoB_2$ are pi-bonding in nature, a much weaker bond than is found in $MgB_2$. The dominance of Mo states at $E_F$ explains the relatively weak dependence of Tc on boron mass that we have observed in our isotope effect measurements.

The calculations indicate that while $MgB_2$ is predominantly a two-dimensional conductor, $MoB_2$ is not. Crystal orbital Hamilton populations (COHPs) are plotted in Figure 16 to illustrate this point. In this type of representation of the electronic states, positive values indicate bonding interactions and negative values indicate antibonding interactions. In this structure, Molybdenum has four nearest like neighbors in the basal ($a$-$b$) plane at a distance of 3.055 Å as well as two out of plane in the adjacent Mo layers at a distance of 3.128 Å. The COHP values for the Mo-Mo in plane and out of plane interactions are both antibonding, and of comparable magnitude. Further, there is a degree of hybridization between B sigma and Mo $d$ orbitals, also previously noted in other transition metal diborides, such as $TaB_2$ (17). Conversely, in $MgB_2$, hybridization between the B-Mg planes or between adjacent Mg planes is almost non-existent, resulting in a two dimensional structure and a boron sublattice that is essentially electronically independent.

The calculated density of states $N(E_F)$ at the Fermi energy for $MoB_2$ is 1.35 per eV cell, and is larger than values reported (17) for $TaB_2$ (0.91 states per eV cell), $ZrB_2$ (0.28), and $NbB_2$ (1.03) with the same structure. As $T_c$ is proportional to $N(E_F)$ in a BCS superconductor, this is likely part of the explanation as to why of these only $MoB_2$ and $NbB_2$ can be made superconducting, the latter at a very low temperature.

Finally, the above calculations allow us to calculate the specific heat coefficient $\gamma$ for $MoB_2$, assuming the free electron model, to be 3.18 mJ/mol $K^2$. The experimentally measured specific heat coefficients for $(Mo_{.96}Zr_{.04})_xB_2$ are 3.65 mJ/mol $K^2$ for x=1 (Tc= 5.9K) and 4.4 mJ/mol $K^2$ for x=0.88 (Tc=8.3K). By comparing the observed and calculated values using the equation $\gamma_{exp} = \gamma_{calc} (1 + \lambda)$, the electron phonon coupling factor ($\lambda$) can be estimated. Thus we estimate that $\lambda \approx$ 0.1 - 0.3 for $(Mo_{.96}Zr_{.04})_xB_2$ (x =1 and x = 0.88, respectively, making the approximation for the latter that the calculated density of states is unchanged from the stoichiometric material), well within the weak coupling limit. For comparison, $\lambda \approx 2$ is obtained for $MgB_2$ when calculated using the same method (19).

## Conclusions

The boron excess molybdenum zirconium diboride superconductor first reported in the 1970s has been shown here to be a Zr-stabilized material of the $AlB_2$ structure type analogous to $MgB_2$, with the boron non-stoichiometry accommodated by the presence of vacant metal sites. The superconducting transition temperature was found to be limited to a maximum of 8.3 K, considerably lower than that of $MgB_2$, and no sign of superconductivity at temperatures above 10K as originally reported was observed. Characterization of the superconductivity and electronic structure calculations reinforce the idea that transition metal based diboride superconductors are electronically quite different from $MgB_2$, with weak electron-phonon coupling and metal derived electronic states dominant at the Fermi energy. If new high temperature superconductors are to be found in the boride family, it appears that the electronic influence of other elements present in the superconductor may have to be minimized.

## Acknowledgements

The work at Princeton University was supported by the NSF, grant DMR98-08941, and the US Department of Energy, grant DE-FG02-98ER45706. The work at Argonne National Laboratory was supported by the US Department of Energy, Office of Basic Energy Science, contract No. W-31-109-ENG-38. The contributions of S. Short to the collection of the neutron diffraction data are acknowledged.

**Figure Captions**

Figure 1. Time-of-flight neutron diffraction pattern of the sample of nominal composition $(Mo_{.96}Zr_{.04})B_{3.1}$. The pattern shows the presence of both the $AlB_2$ type superconducting compound $(Mo_{.96}Zr_{.04})_{.85}B_2$ (peaks shown by lower row of tic marks) and elemental boron (upper row of tic marks). Difference between observed and calculated intensities shown under the diffraction pattern. Inset shows the region of the pattern with the strongest boron peaks.

Figure 2. Variation of the composition of the $AlB_2$ type phase of formula $(Mo_{.96}Zr_{.04})_xB_2$ as a function of the as-prepared boron to metal ratio, B/M.

Figure 3. The experimentally determined boron to metal ratio, including the boron in the diboride phase and that present as second phase elemental born, as a function of the as-prepared nominal boron to metal ratio. The 1:1 correspondence confirms the validity of the structural and chemical model.

Figure 4. The variation of the in-plane (a) and out-of-plane (c) lattice parameters as a function of stoichiometry of the $AlB_2$ type phase for $(Mo_{.96}Zr_{.04})_xB_2$.

Figure 5. The [100] electron diffraction pattern for stoichiometric $(Mo_{.96}Zr_{.04})B_2$. The diffraction pattern illustrates the presence of streaks along the c axis, indicative of structural disorder in the direction perpendicular to the planes.

Figure 6a. [100] low resolution electron microscope image showing the presence of significant numbers of planar defects perpendicular to the c axis in stoichiometric $(Mo_{.96}Zr_{.04})B_2$. Not all planar defects are clearly visible in this image: well imaged defect planes indicated by black arrows. The general light and dark contrast in the image indicates the presence of significant strain throughout the whole structure induced by the planar defects.

Figure 6b: Higher resolution image of the stacking defects, indicated by black arrows. The inset is a high resolution image enlarging the upper defect, averaged over 4 unit cells along the horizontal direction. The white bars in the inset emphasize the *c*-axis spacing difference of the defect.

Figure 7. Electron diffraction pattern in the [011] reciprocal lattice plane of the metal deficient diboride $(Mo_{.96}Zr_{.04})_{.88}B_2$. The very weak superlatice reflections are shown.

Figure 8. Magnetic characterization of the superconducting transitions as a function of composition for the diboride superconductor $(Mo_{.96}Zr_{.04})_xB_2$.

Figure 9. Variation of superconducting transition temperature with stoichiometry for $(Mo_{.96}Zr_{.04})_xB_2$. Inset: variation of Tc with crystallographic cell parameter ratio c/a.

Figure 10. Characterization of the resistivity of $(Mo_{.96}Zr_{.04})_{.88}B_2$ in the normal state, and in the vicinity of Tc (inset).

Figure 11. Characterization of the superconducting transitions of stoichiometric $(Mo_{.96}Zr_{.04})B_2$ and nonstoichiometric $(Mo_{.96}Zr_{.04})B_2$ diboride superconductors by specific

heat. Dotted lines in the lower panel indicate the extrapolation to $T^2 = 0$ employed to determine the electronic density of states.

Figure 12. Detailed comparison of the specific heats in the vicinity of Tc for $(Mo_{.96}Zr_{.04})_{.88}B_2$ made with $^{10}B$ and $^{11}B$.

Figure 13. Magnetic characterization of the superconducting transitions for molybdenum-based diboride phases with the nominal compositions shown.

Figure 14. Magnetic characterization of the superconducting transitions for niobium-based diboride phases with the nominal compositions shown.

Figure 15. Electronic density of states (DOS) versus energy for $MoB_2$, in hexagonal space group P6/mmm. Total DOS as well as Mo and B partial DOS are plotted. States at the Fermi energy ($E_F$) are roughly 75% Mo in character.

Figure 16. Negative crystal orbital Hamilton population (-COHP) versus energy for nearest neighbor interactions. Positive –COHP values indicate bonding interactions, while negative values indicate antibonding interactions. Energy is plotted relative to the Fermi level ($E_F$).

**Table 1: Structural Parameters for $(Mo_{.96}Zr_{.04})_xB_2$ Determined by Neutron Diffraction.** $AlB_2$ type structure, hexagonal symmetry, space group P6/mmm. $(Mo_{.96}Zr_{.04})$ in site 1a, (0,0,0), B in site 2d, (1/3, 2/3, 1/2 )*

| B/M | a (Å) | c (Å) | c/a | Vol. (Å³) | Occ(M) | $U_i$(M) | U11(B) | U33(B) | wt%(B) | B/M$_{exp}$ |
|---|---|---|---|---|---|---|---|---|---|---|
| 2.0 | 3.05517(6) | 3.12813(16) | 1.02388(7) | 25.286(41) | 1.003(16) | 0.34(4) | 0.40(4) | 2.12(18) | 0 | 1.99(6) |
| 2.1 | 3.05222(5) | 3.14160(12) | 1.02928(6) | 25.346(32) | 0.961(13) | 0.35(4) | 0.41(4) | 1.88(12) | 0 | 2.08(6) |
| 2.2 | 3.04932(4) | 3.15570(9) | 1.03489(4) | 25.412(25) | 0.908(10) | 0.47(3) | 0.47(3) | 1.84(9) | 0 | 2.20(5) |
| 2.3 | 3.04707(3) | 3.15975(7) | 1.03698(3) | 25.407(19) | 0.882(7) | 0.49(3) | 0.52(2) | 1.45(6) | 0.58(8) | 2.33(5) |
| 2.4 | 3.04656(3) | 3.16078(7) | 1.03749(3) | 25.406(19) | 0.852(7) | 0.49(3) | 0.54(2) | 1.65(7) | 1.17(8) | 2.48(5) |
| 2.5 | 3.04606(3) | 3.15978(6) | 1.03733(3) | 25.390(18) | 0.865(6) | 0.47(2) | 0.49(2) | 1.40(5) | 1.79(7) | 2.51(5) |
| 2.6 | 3.04397(3) | 3.16067(6) | 1.03834(3) | 25.362(18) | 0.843(7) | 0.65(2) | 0.70(2) | 1.57(5) | 2.79(9) | 2.69(6) |
| 2.7 | 3.04475(3) | 3.15872(7) | 1.03743(3) | 25.360(19) | 0.843(7) | 0.52(3) | 0.56(2) | 1.49(6) | 4.04(11) | 2.84(7) |
| 2.8 | 3.04520(3) | 3.15941(5) | 1.03750(3) | 25.373(16) | 0.844(6) | 0.48(2) | 0.53(2) | 1.45(5) | 3.28(9) | 2.75(5) |
| 2.9 | 3.04430(3) | 3.15764(6) | 1.03723(3) | 25.344(18) | 0.847(7) | 0.49(3) | 0.56(2) | 1.31(5) | 4.98(13) | 2.94(7) |

* B/M, nominal boron to metal ratio; *a,c* refined cell parameters; Occ(M) refined occupancy of the metal site; $U_i$(M) isotropic thermal parameter for metal atom; U11(B), U33(B), anisotropic thermal parameters for boron; wt%(B), refined weight percent boron second phase; B/M experimentally determined boron to metal ratio in sample.

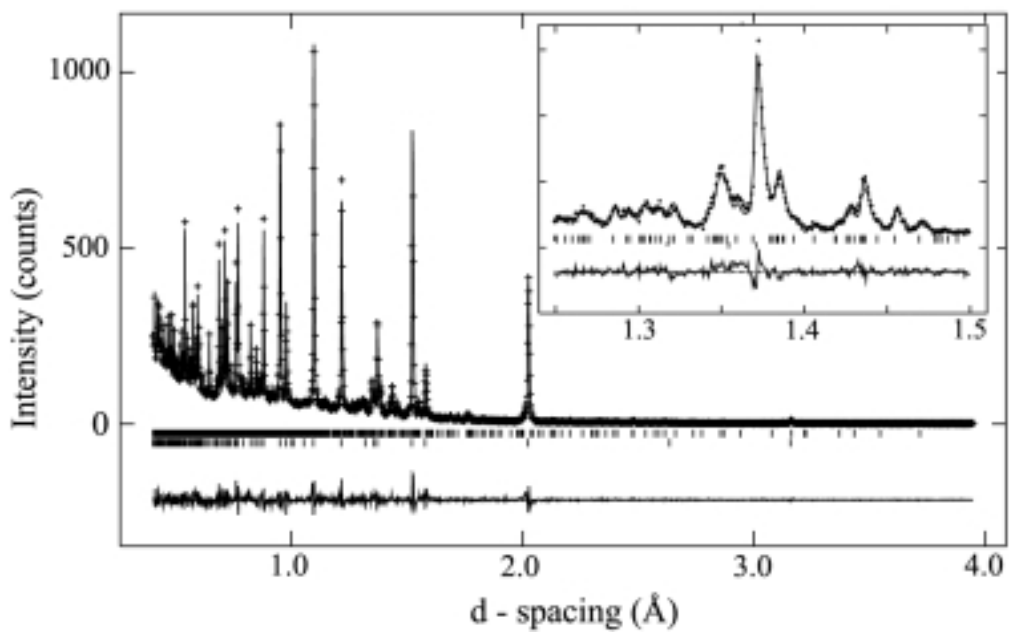

**Figure 1**

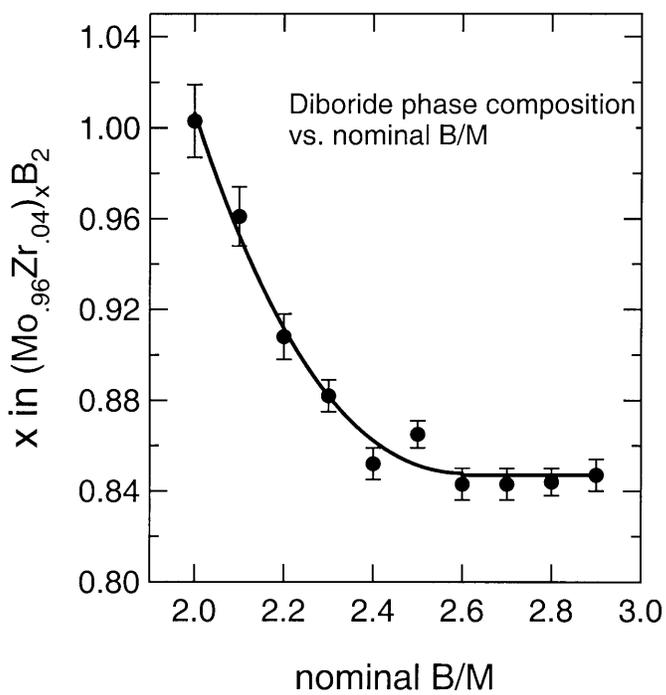

**Figure 2**

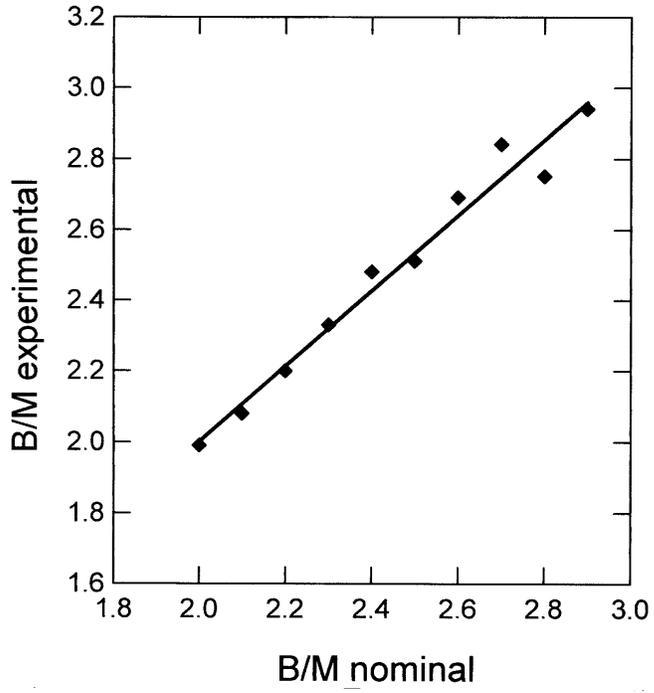

Figure 3

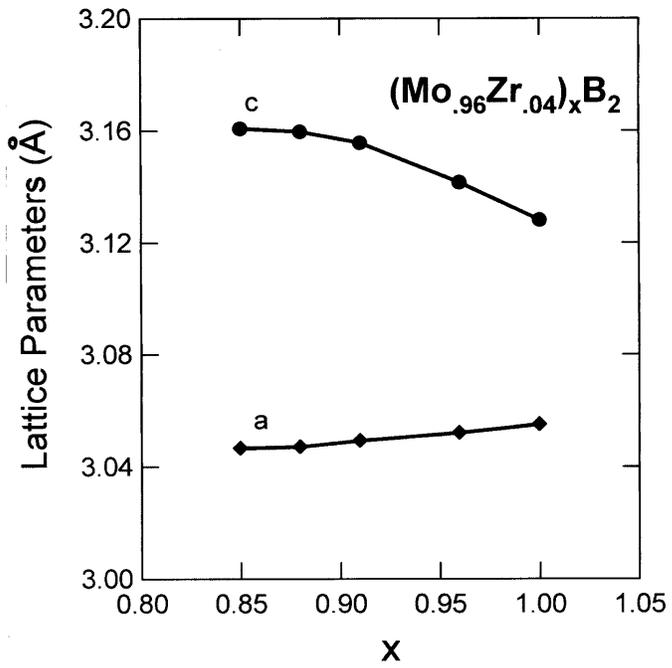

Figure 4

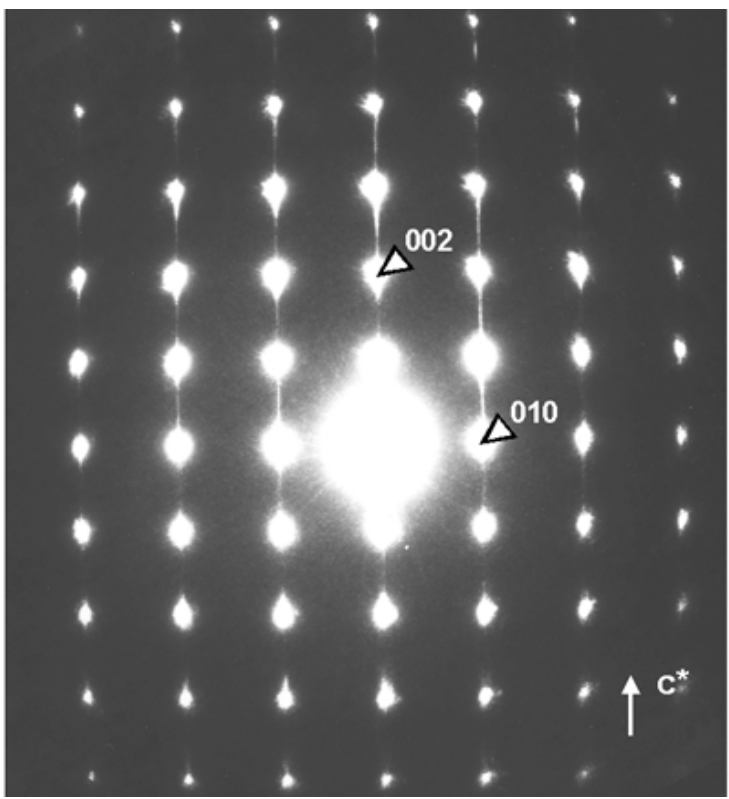

**Figure 5**

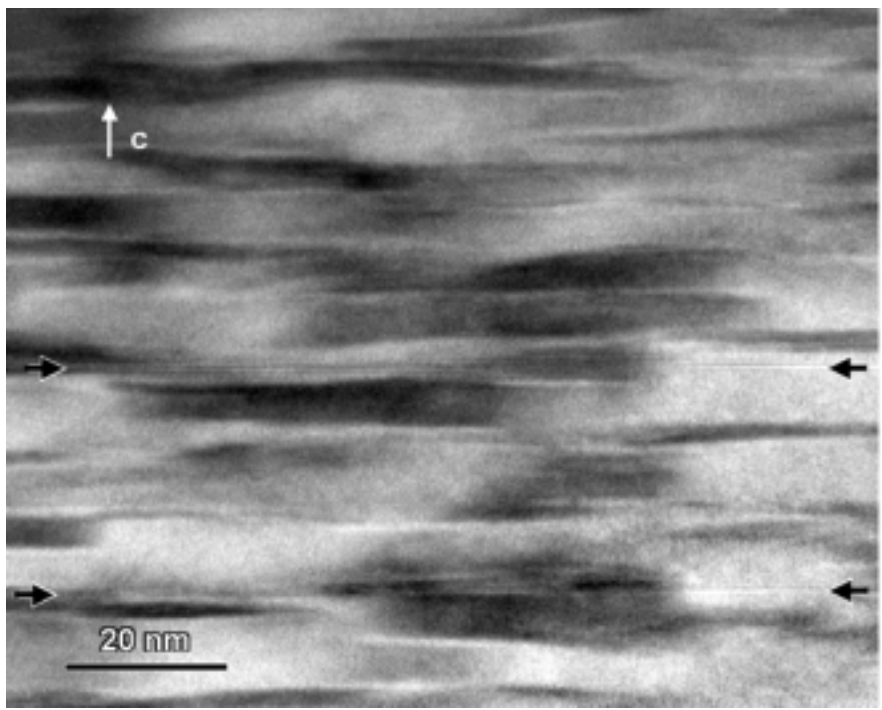

**Figure 6a**

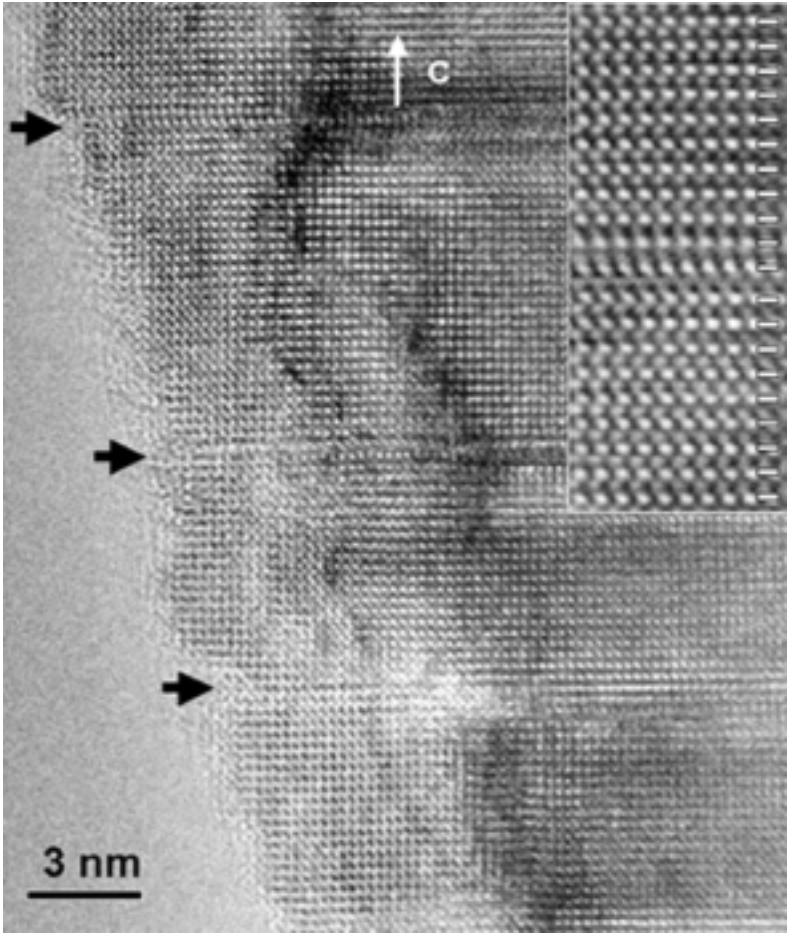

**Figure 6b**

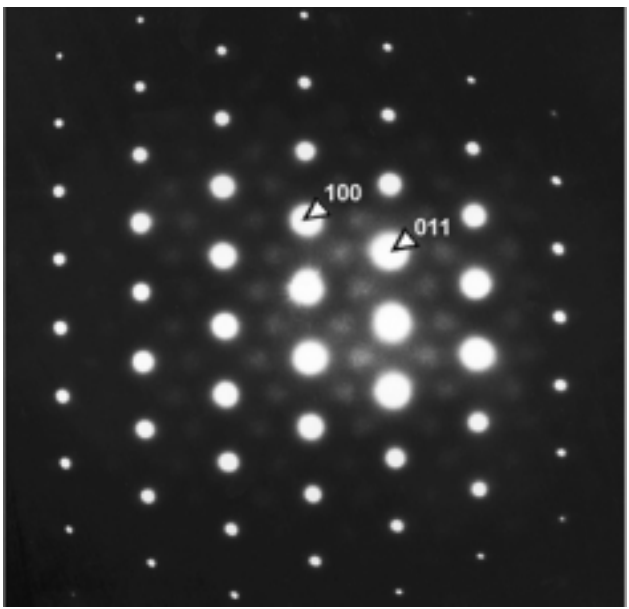

**Figure 7**

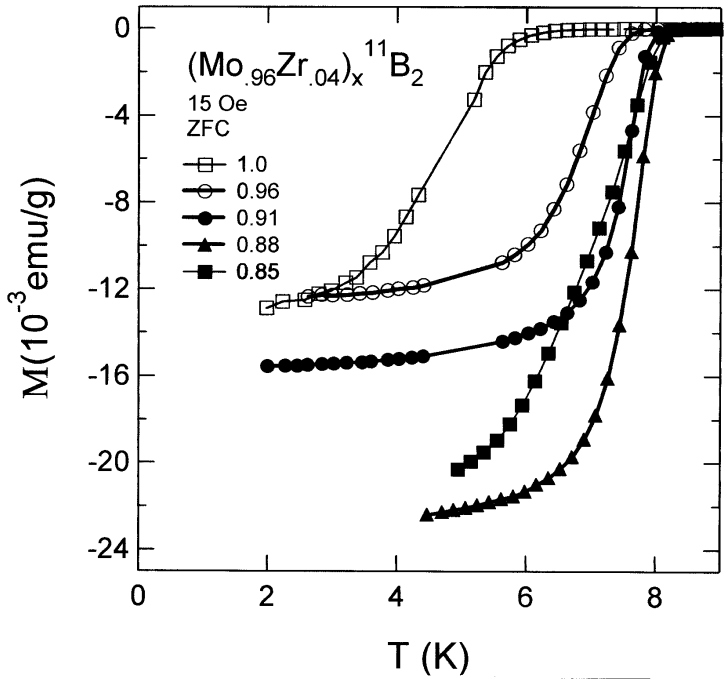

Figure 8

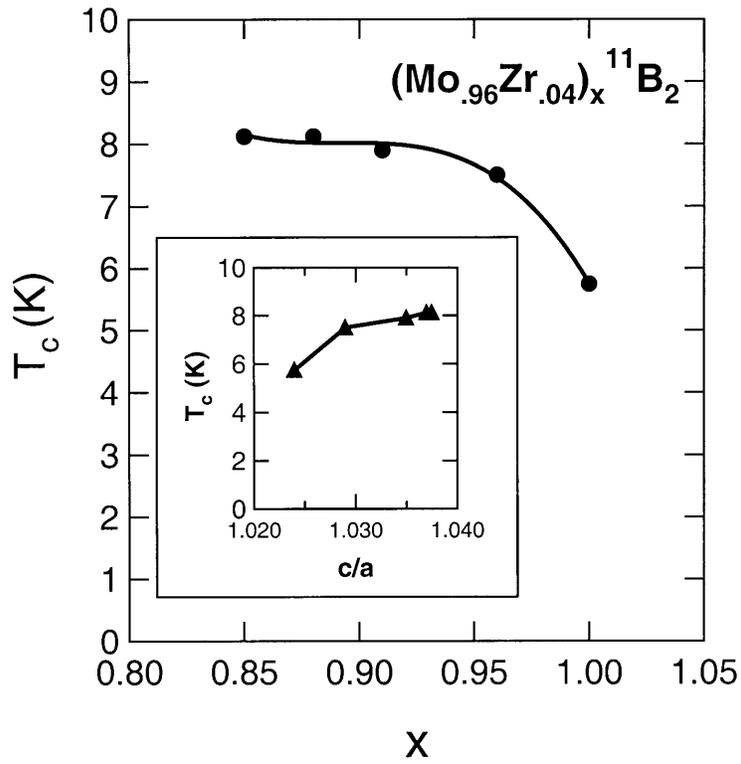

Figure 9

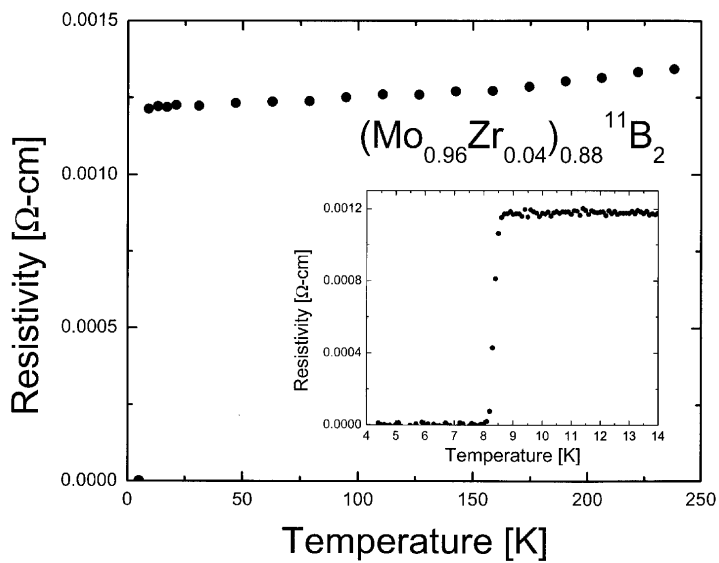

Figure 10

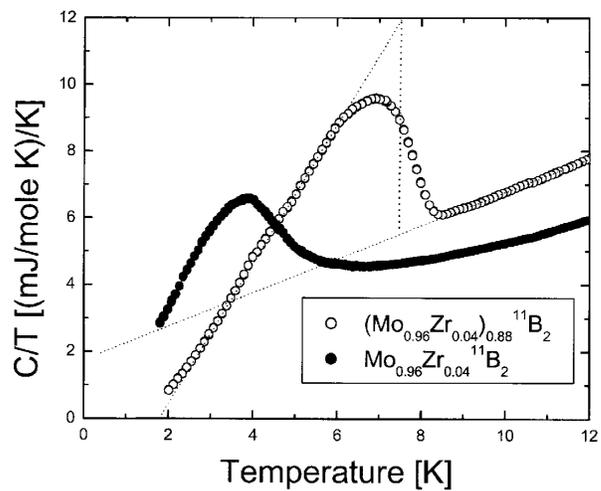

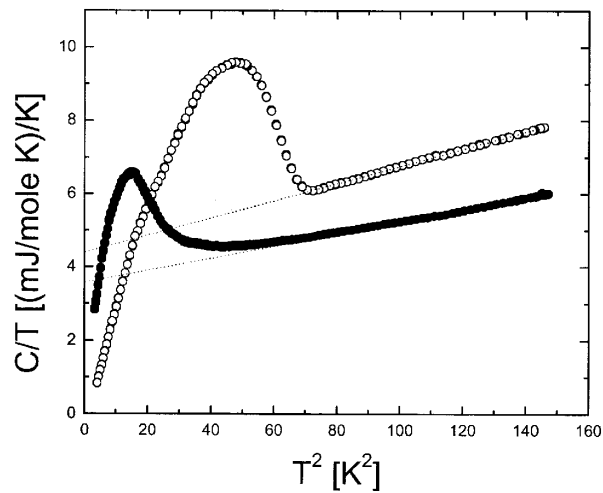

Figure 11

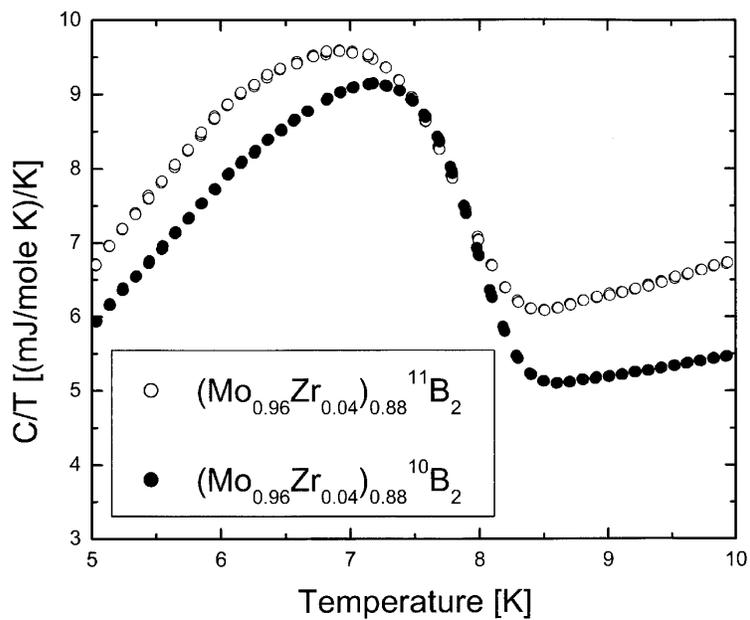

Figure 12

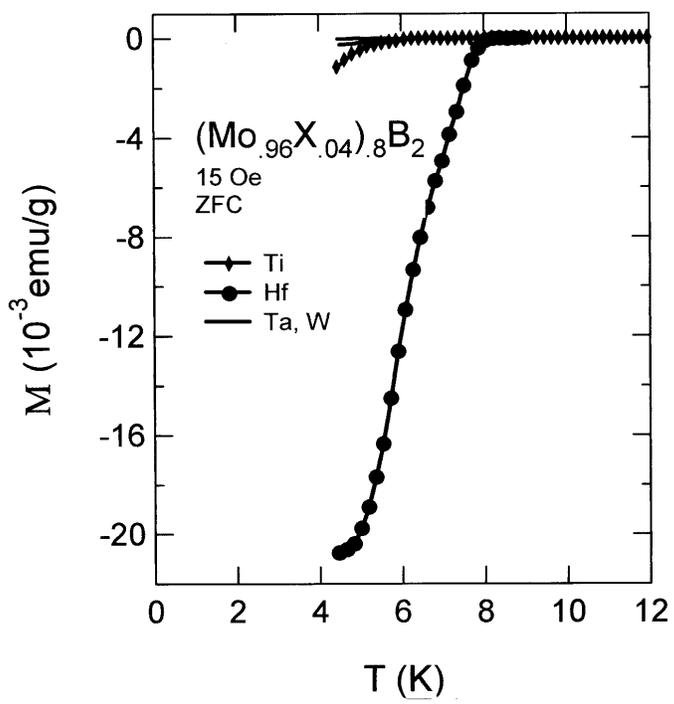

Figure 13

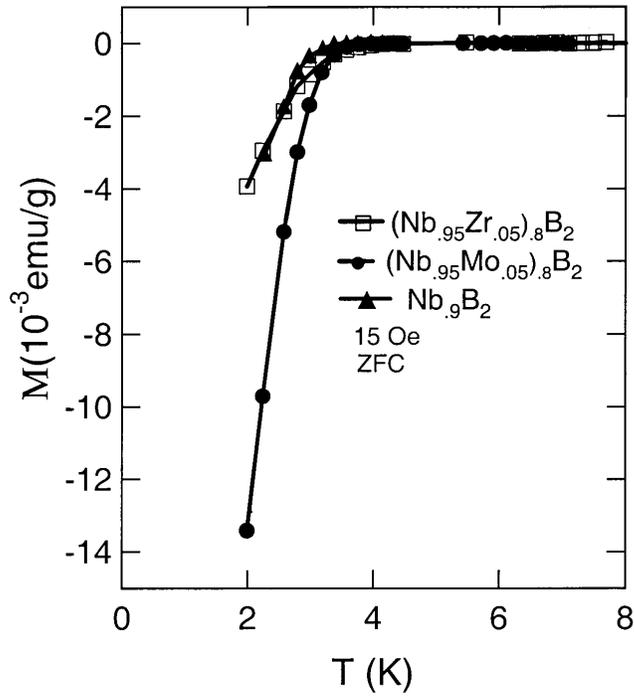

Figure 14

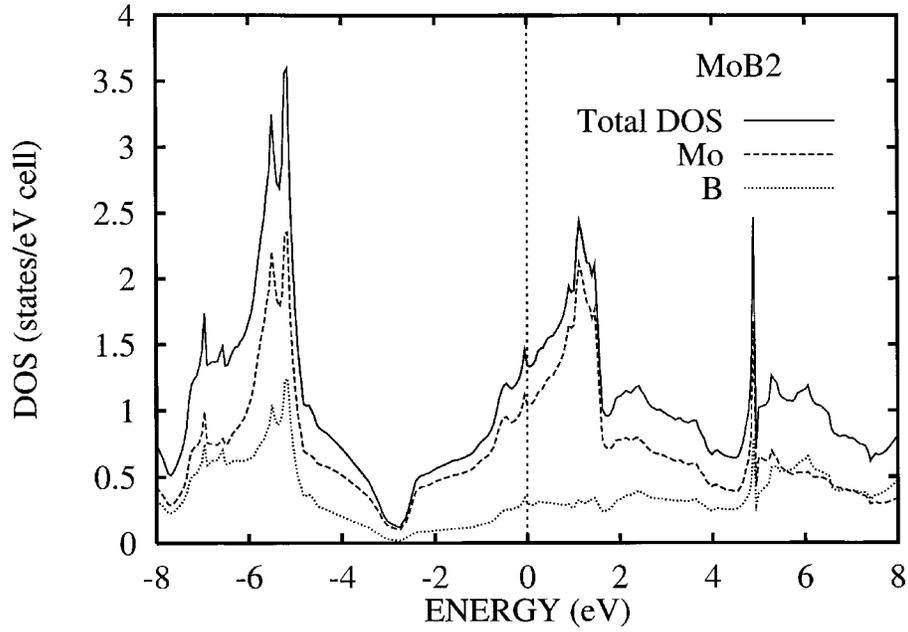

Figure 15

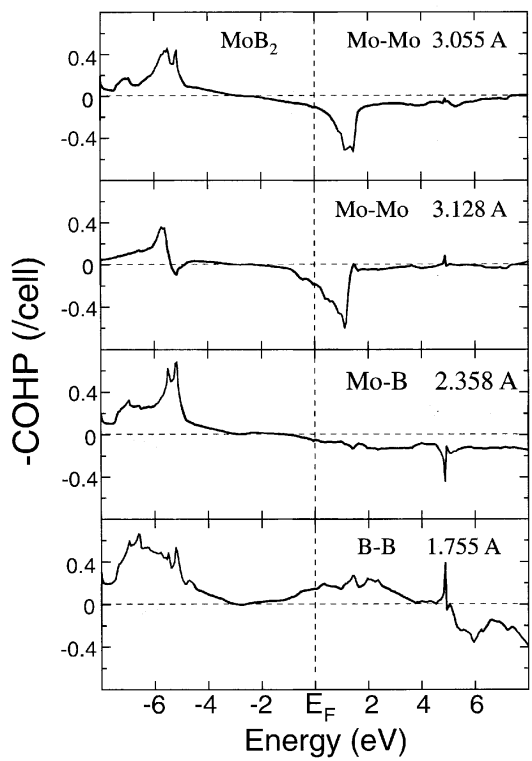

**Figure 16**